\documentclass[modern]{aastex61}

\usepackage{graphicx}% Including figure files
\usepackage{amsmath}% Advanced maths commands
\usepackage{amssymb}% Extra maths symbols
\usepackage{enumitem}

\newcommand\mybar{\kern1pt\rule[-\dp\strutbox]{.8pt}{\baselineskip}\kern1pt}

\setlist[itemize]{noitemsep, topsep=0pt, leftmargin=*}

\shorttitle{Surface Brightness Bias}
\shortauthors{Loeb}

\begin{document}

\title{Surface Brightness Bias in the Shape Statistics of High-Redshift Galaxies}

\author{Abraham Loeb}
\affiliation{Astronomy Department, Harvard University, 60 Garden
  St., Cambridge, MA 02138, USA}

\begin{abstract}
Recently, Pandya et al. (2023) argued that the shapes of dwarf
galaxies in JWST-CEERS observations show a prolate fraction that rises
from $\sim 25\%$ at redshifts $z=0.5$-$1$ to $\sim 50$-$80\%$ at
$z=3$-$8$. Here we suggest that this apparent change could result from
a surface-brightness bias, favoring the detection of edge-on disks at
low-luminosities and high redshifts. Changing edge-on projections with
an axis ratio of 10 to a face-on orientation reduces their apparent
surface brightness by 2.5 magnitude per arcsec$^2$ and could shift a
substantial fraction of the observed galaxies below the detection
limit.
\end{abstract}

\section{Introduction}

A recent statistical analysis of the apparent shapes of dwarf galaxies
with stellar masses in the range ${\rm log_{10}} (M_\star/M_\odot)=
9-10.5$ at redshifts $z=0.5$-$8$ in JWST-CEERS
observations~\citep{Fink}, concluded that prolate fraction rises at
low stellar masses from $\sim 25\%$ at redshifts $z=0.5$-$1$ to $\sim
50$-$80\%$ at $z=3$-$8$~\citep{Pandya}.  Here we suggest that this
change might reflect a bias as a result of the requirement that the
galaxies exceed a surface brightness threshold for detectability
against the sky background noise.

\section{Method}

The projected surface brightness of a circular galactic disk which is
tilted by an angle $\alpha$ relative the line-of-sight would be
enhanced by a factor of $\sim 1/\sin\alpha$ relative to a face-on
orientation with $\alpha=\pi/2$. This factor also equals the axis
ratio $(a/b)$ for the projected ellispoidal image, which increases the
face-on value of $(a/b)=1$ by this factor. The observed surface
brightness should also decline with increasing redshift as
$(1+z)^{-4}$~\citep{LF13} at a fixed rest-frame wavelength band.

For a given stellar mass $M_\star$, the sizes of galaxies with ${\rm
  log_{10}} (M_\star/M_\odot)=10.5$ in CEERS were observed to decline
with increasing redshifts as $(1+z)^{-0.63}$~\citep{Ward}. When
combined with the $(1+z)^{-4}$ dimming, this scaling implies that at a
fixed $M_\star$ and rest-frame wavelength band, the surface brightness
(observed radiation energy in that band per unit time per unit angular
area on the sky) of projected galaxy images scales as,
\begin{equation}
\mu \propto {(a/b)\over (1+z)^{2.74}}.
\label{eq:scale}
\end{equation}

In addition, the size-luminosity relation of galaxies in the CEERS
survey shows a logaithmic slope in the rest-frame optical of $\sim
0.2$~(see Fig. 6 in~\citet{Ward}). Adopting a universal mass-to-light
ratio (or equivalently, initial mass function) for the stars, yields
the additional scaling with $M_\star$ of,
\begin{equation}
\mu \propto {M_\star^{0.6} (a/b)\over (1+z)^{2.74}}.
\label{eq:scale}
\end{equation}

The emission from low-luminosity galaxies is detected only out to a
surface-brightness contour that exceeds the sensitivity threshold
relative to the background noise. For dwarf galaxies near the threshold of
detectability, one expects an enhanced occurence rate of $(a/b)>1$, as
inferred by \citet{Pandya} at high-$z$ and low-$M_\star$ values.

\section{Results}

At a fixed stellar mass bin in Table 2 of \citet{Pandya}, the
surface-brightness bias implies a scaling of detectability threshold
for $(a/b)$ with $(1+z)^{2.74}$. The transition between the redshift
intervals $z= 0.75\pm 0.25; 1.25 \pm 0.25; 1.75\pm 0.25; 2.25\pm 0.25;
2.75\pm 0.25;$ and $z>3$, translate to biases of $(a/b)$ by factors of
$1; 1.99; 3.45; 5.45; 8.07$ and $>9.63$, respectively. These factors
are moderated at higher galaxy luminosities by the above-mentioned
scaling of $\mu$ with $M_\star^{0.6}$, implying that galaxies in the
mass bin of ${\rm log_{10}} (M_\star/M_\odot) = 10$-$10.5$ are a
factor of $\sim 3.98$ brighter in surface brightness above the
$\mu$-detectability threshold, when compared to fainter galaxies in
the mass bin of ${\rm log_{10}} (M_\star/M_\odot) = 9$-$9.5$. If the
low-mass bin shows a bias at $z>5$, then the high-mass bin will show
this bias only at $z>8.93$, above the observed range of redshifts in
Table 2 of~\citet{Pandya}. This result could explain the reported
prevalence of prolate galaxies in the JWST-CEERS images at
low-$M_\star$ and high-$z$ values.

The elongated galaxies identified by \cite{Pandya} show axis ratios of
up to $(a/b)\sim 10$.  Changing edge-on projections with an axis ratio
of 10 to a face-on orientation, reduces their apparent surface
brightness by 2.5 magnitude per arcsec$^2$. Given that the low-mass,
high-redshift galaxy sample with ${\rm log_{10}} (M_\star/M_\odot) =
9$-$9.5$ at $z=3$-8 is characterized by a surface brightness of
$27~{\rm AB~mag~arcsec}^{-2}$, a reduction by 2.5 AB magnitudes could
shift a substantial fraction of the elongated galaxies below the
detection limit.

Theoretically, \citet{Costantin23} used simulated galaxy images to
show that JWST-CEERS images are expected to miss the outer, low
surface-brightness parts of high-redshift galaxies.

\section{Discussion}

The numbers derived above indicate that the reported statistical
prevalence of prolate shapes in the mass bin of ${\rm log_{10}}
(M_\star/M_\odot) = 9$-$9.5$ of galaxies at redshifts $z=3$-$8$ could
reflect a surface-brightness bias and not an intrinsic property of
high-redshift galaxies. This JWST-CEERS sample of galaxies was argued
to be complete in Appendix B of~\citet{Pandya}, but this analysis was
framed in terms of luminosity (magnitude) and not surface brightness,
which affects the inference based on simulations of high-redshift
galaxies~\citep{Costantin23}.

Of course, there are also astrophysical explanations at play, such as
an enhanced merger fraction at increasing redshift, which would lead
to elongated tidal tails~\citep{Dom}. Indeed, \citet{Whitney20} found
strong redshift evolution in the surface brightness of galaxies based
on data from the Hubble Space Telescope, consistent with evolution in
merger rate. Additional contributions might originate from shear
associated with weak gravitational lensing by foreground galaxies or
enhanced star formation in galactic bars.  However, the
surface-brightness bias discussed here must be corrected for before
any astrophysical conclusions are drawn on the basis of raw JWST-CEERS
images.

\bigskip
\bigskip
\section*{Acknowledgements}

This work was supported in part by Harvard's {\it Black Hole
  Initiative}, which is funded by grants from JFT and GBMF.  

\bigskip
\bigskip
\bigskip

\bibliographystyle{aasjournal}
\bibliography{t}
\label{lastpage}
\end{document}